\documentclass[twocolumn,showpacs,prl,amsmath,amssymb]{revtex4}

\usepackage{graphicx}
\usepackage{dcolumn}
\usepackage{bm}
\usepackage{epsfig} 

\def\eq#1{{Eq.~(\ref{#1})}}

\def\vev#1{\left\langle #1\right\rangle}

\def\hbar{\hspace{0pt}\raisebox{1pt}{$-$} \hspace{-7pt} h}

\def\5{\overline 5}

\newcommand{\be}{\begin{equation}}
\newcommand{\ee}{\end{equation}}
\newcommand{\bea}{\begin{eqnarray}}
\newcommand{\eea}{\end{eqnarray}}
\newcommand{\nn}{\nonumber}

\let\vev\VEV
\def\baselinestretch{1.4}
\def\roughly#1{\mathrel{\raise.3ex\hbox{$#1$\kern-.75em
      \lower1ex\hbox{$\sim$}}}} 

\def\e6{E(6)}
\def\321{$SU(3)_{c}\otimes SU(2)_L \otimes U(1)$}
\def\10{SO(10)}

\def\422{SU(4) $\otimes$ SU(2) $\otimes$ SU(2)}

\newcommand{\AHEP}{Instituto de F\'{\i}sica Corpuscular --
  C.S.I.C./Universitat de Val{\`e}ncia \\
  Campus de Paterna, Apt 22085,
  E--46071 Val{\`e}ncia, Spain}

\begin{document}
\preprint{IFIC/06-24}

\title{Neutrino masses and electroweak symmetry breaking}

\author{F. Bazzocchi}
\affiliation{\AHEP}
\author{J. W. F. Valle}
\email{valle@ific.uv.es}
\homepage{http://ahep.uv.es}
\affiliation{\AHEP}

\date{\today}

\begin{abstract} 

  Neutrino mass generation may affect the basic structure of the
  electroweak symmetry breaking sector.  We consider a broad class of
  elementary particle theories where neutrinos get mass at a low mass
  scale.
  We show how these can be made natural up to few TeV or so, in the
  absence of supersymmetry or other possible stabilizing mechanisms.
  Although the standard signatures for which LHC has been optimized
  are absent, others are expected. A generic one among these is the
  possibility of an invisibly decaying Higgs boson which is
  characteristic of models with spontaneous breaking of lepton number
  symmetry below TeV or so.

\end{abstract}

\maketitle



Two important theoretical challenges of particle physics are
elucidating the nature of electroweak symmetry breaking and the origin
of neutrino masses. While the neutrino oscillation data confirmed the
existence of neutrino masses and mixing~\cite{Maltoni:2004ei}, the
Higgs boson has not yet given an irrefutable proof of its existence.

It has been suggested by Barbieri and Hall~\cite{Barbieri:2005kf} that
the physics that tames the quadratic Higgs boson mass divergence may
be much less ambitious than supersymmetry.
Here we follow up on this idea and propose that the physics
responsible for ``postponing'' naturalness is the same one providing
neutrinos their mass.

We point out that neutrino masses, however small, may drastically
affect the electroweak sector. While some of our considerations hold
for generic models with explicitly broken lepton number, we will focus
on models with spontaneous violation of lepton
number. 
In such models a lepton number symmetry is broken by an \321 singlet
vacuum expectation value $\vev{\sigma} $ below a few
TeV~\cite{Joshipura:1992hp}.
It has long been noted that in all these models the Higgs has the
"invisible" mode
\begin{equation}
  \label{eq:jj}
  h \to JJ
\end{equation}
as a sizeable decay channel~\cite{Joshipura:1992hp}, which may
dominate over the SM modes, such as $b\bar b$ or $\tau\bar \tau$.
Here $J$ denotes the associated pseudoscalar Goldstone boson, called
majoron. Since it is weakly interacting with all other particles, this
leads to events with large missing energy that could be observable at
collider
experiments~\cite{Joshipura:1992hp,deCampos:1997bg,Abdallah:2003ry}.

Here we note that, apart from changing the low energy theory by the
existence of the new Higgs decay in Eq.~(\ref{eq:jj}) such models
generically improve the naturalness of the electroweak symmetry
breaking sector even in the absence of supersymmetry or some other
kind of physics capable of canceling the quadratically divergent top
loop contribution to the Higgs boson mass.  Therefore they provide a
``worse-case'' scenario for LHC~\cite{Barbieri:2005kf} in which
naturalness in the electroweak symmetry breaking sector is effectively
``lifted'' to a scale higher than testable at the LHC. 

However, the novel properties of the scalar sector itself could give
visible ``signs of life'' in collider experiments, for example, the
mode in Eq.~(\ref{eq:jj}) and many other signatures which depend on
how neutrinos get their masses by coupling to the scalars.

Consider the simplest tree level scalar potential that can
simultaneously account for electroweak symmetry breaking and the
generation of naturally small neutrino masses,
\begin{eqnarray}
\label{eq:pot}
V [\Phi,\sigma]= \mu^2_{0_{\Phi}} (\Phi^\dag \Phi) +\mu^2_{0_{\sigma}} (\sigma^\dag \sigma) + \nn \\
\lambda_1  (\Phi^\dag \Phi)^2+ \lambda_\sigma  (\sigma^\dag \sigma)^2+\lambda_{\Phi \sigma} (\Phi^\dag \Phi)(\sigma^\dag \sigma)
\end{eqnarray}
The potential in Eq.~(\ref{eq:pot}) has a U(1) global symmetry under
which $ \sigma$ gets rephased, but not $ \Phi$. This will become
lepton number once leptons are coupled (see example below).

Once we include the one-loop quadratic divergent corrections to the
scalar potential the previous $\mu^2_{0_{i}}$ are replaced by
\begin{eqnarray}
\label{eq:shiftmu}
\mu^2_\Phi= \mu^2_{0_{\Phi}}  + \frac{\Lambda^2}{16 \pi^2} \gamma_\Phi \nn\\
\mu^2_\sigma= \mu^2_{0_{\sigma}}  + \frac{\Lambda^2}{16 \pi^2} \gamma_\sigma \,,
\end{eqnarray}
where $\gamma_{\Phi}$ takes into account the gauge, the top and scalar
contributions and $\gamma_{\sigma}$ only the scalar one.  The
parametrization given by Eq.~(\ref{eq:shiftmu}) allows us to discuss
the stabilization of the vevs in terms of their dependence on the cut
off scale $\Lambda^2$, i.~e., in terms of the fine tuning.

The vacuum configuration that breaks the electroweak gauge symmetry
and the $U(1)$ global symmetry through
\[
\vev{\Phi}= v_\Phi/\sqrt{2}~~~~ \vev{\sigma}= v_\sigma/\sqrt{2} \,.
\]

By solving the extremization conditions we can write the expressions
of the vevs in terms of the parameters of the potential
\begin{eqnarray}
\label{eq:vevexp}
v_\Phi^2 &=&2\frac{\lambda_{\Phi \sigma} \mu_\sigma^2 -2 \lambda_\sigma \mu_\Phi^2}{4 \lambda_\Phi \lambda_\sigma -\lambda_{\Phi \sigma}^2} \nn\\
v_\sigma^2 &=& 2\frac{\lambda_{\Phi \sigma} \mu_\Phi^2 -2 \lambda_\Phi \mu_\sigma^2}{4 \lambda_\Phi \lambda_\sigma -\lambda_{\Phi \sigma}^2} 
\end{eqnarray}

Following \cite{Casas:2004gh} we can define the fine tuning parameter
\begin{equation}
\label{eq:ft}
D=\sqrt{D_{v_\Phi}^2+D_{v_\Phi}^2}\,.
\end{equation}
where $ D_{v_\Phi}= \frac{\partial \ln v_\Phi^2}{\partial \ln
  \Lambda^2}$ and $ D_{v_\sigma}= \frac{\partial \ln
  v_\sigma^2}{\partial \ln \Lambda^2}$.  Inserting these in
Eq.~(\ref{eq:ft}) and inverting the expression we obtain an upper
bound for the cut-off 
\begin{equation}
\label{eq:Lambda}
\Lambda^2 \leq  8 \pi^2 v^2_\Phi v^2_\sigma D F
\end{equation}
where $F$ is given by
\begin{eqnarray}
\frac{( 4 \lambda_\Phi \lambda_\sigma- \lambda_{\Phi\sigma}^2)}{\sqrt{(v_\Phi^4(-2 \gamma_\sigma \lambda_\Phi + \gamma_\Phi \lambda_{\Phi \sigma})^2+ v_\sigma^4(-2 \gamma_\Phi \lambda_\sigma + \gamma_\sigma \lambda_{\Phi \sigma})^2}}\,. \nn
\end{eqnarray}
Minimizing the potential one obtains the mass matrix
\begin{eqnarray}
M= \left( \begin{array} {cc}
2 \lambda_\Phi v_\Phi^2 &\lambda_{\Phi \sigma} v_\Phi v_\sigma \\
\lambda_{\Phi \sigma} v_\Phi v_\sigma &2 \lambda_\sigma v_\sigma^2
\end{array} \right) \,,
\end{eqnarray}
for the two neutral CP-even scalars, $h$ and $H$.
Its diagonalization leads to a mixing angle between doublet and
singlet states in the CP-even sector, given by
\begin{equation}
  \label{eq:alfa}
\tan 2\alpha =
\lambda_{\Phi \sigma} v_\Phi v_\sigma/(\lambda_\sigma v_\sigma^2-\lambda_\Phi v_\Phi^2)
\end{equation}
and the masses $m^2_{H,h}$ of the two scalars,
\begin{equation}
  \label{eq:masses}
 \lambda_\Phi v_\Phi^2 + \lambda_\sigma v_\sigma^2 \pm \sqrt{ \lambda_\Phi v_\Phi^4+ \lambda_\sigma v_\sigma^4 + v_\Phi^2 v_\sigma^2 (\lambda_{\Phi \sigma}^2-2 \lambda_\Phi \lambda_\sigma)},
\end{equation}
where $m^2_H$ is associated to the plus sign, $m^2_h$ to the minus. In
this scheme the vacuum breaks both electroweak symmetry and lepton
number. Since the latter is a $U(1)$ global symmetry there is, in
addition, a pseudoscalar Goldstone boson, the majoron, to complete the
set of physical spin-less bosons.

\begin{figure}[ht] \centering
 \includegraphics[clip,height=8cm,width=\linewidth]{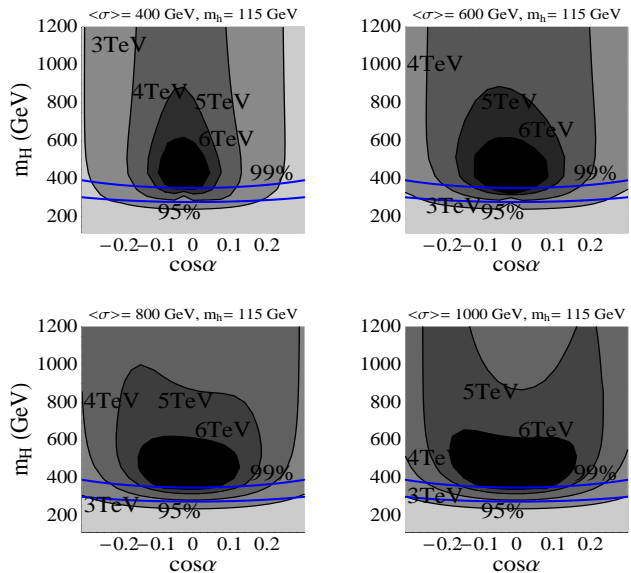}
\caption{\label{fig:cutoff} 
Regions in the $m_H$-$\cos\alpha$ plane leading to an increase
in the cutoff $\Lambda$ for 10\% fine-tuning and $m_h=115$ GeV. In the darkest 
contour $\Lambda\geq6$~TeV and in the lightest $\Lambda\geq2$~TeV, 
decreasing in TeV steps in between. Note that in our model we can 
have the lightest scalar boson with a 
mass of 115 GeV and a cutoff around 6 TeV, while in the SM
for $ m_h=115$ GeV the cutoff is  1.2 TeV. }
\end{figure}

We can use Eqs.~(\ref{eq:alfa})--(\ref{eq:masses})  to rewrite   Eq.~(\ref{eq:Lambda}), to map the
restriction on the cutoff  in terms of the five
parameters $m_{h} , m_H , v_\Phi , v_{\sigma}$ and $\alpha$.
Before doing this let us consider two limiting cases. When $\cos
\alpha=1$ Eq.~(\ref{eq:Lambda}) reduces essentially to the SM case,
\begin{equation}
  \label{eq:limit-1}
\Lambda^2 \leq D  \pi^2 m^2 _h\,,  
\end{equation}
For the case $\cos \alpha=0$ Eq.~(\ref{eq:Lambda}) becomes
\begin{equation}
  \label{eq:limit-2}
\Lambda^2 \leq D \,4 \pi^2 m_H^2 f(x)\,,  
\end{equation}
where 
\begin{eqnarray}
f(x)&=&\left\{ \begin{array}{c}\frac{x^2}{\sqrt{5-24 x^2 +36 x^4}}  \quad v_\Phi \simeq v_\sigma \ll  m_H ,\quad x= v_\Phi/ v_\sigma \\ \frac{x^2}{\sqrt{4-24 x^2 +37 x^4}}  \quad v_\Phi \ll v_\sigma \simeq m_H, \quad x= v_\Phi/ m_H \\
\frac{x^2}{\sqrt{1+16 x^4}}  \quad v_\Phi\simeq m_H \ll v_\sigma, \quad x=  m_H/v_\sigma
\end{array} \right.\nn
\end{eqnarray}
and the cutoff $\Lambda$, being proportional to the heaviest CP-even
scalar,  can be raised up to few TeV for natural choices of the parameters  that give $f(x) \sim O(1)$.
In Fig.~\ref{fig:cutoff} we show the regions of ``extended
naturalness'' is this simplest model. We display the contour regions
in the $m_H$-$ \cos \alpha$ plane leading to an increase in the
effective cutoff $\Lambda$, which can reach a few TeV or so for
reasonable values of parameters in this model.

The parameters of the electroweak symmetry breaking sector are
constrained by electroweak precision tests (EWPT) which indicate a
value for the Higgs boson mass $129^{+74}_{-49}$
GeV~\cite{unknown:2005em} which in our model corresponds to
\begin{equation} \label{eq:condT}
\cos^2 \alpha \log{m_h}+ \sin^2 \alpha \log{m_H} \leq
  \log{m_h^{+2(3)\sigma}} ,   
\end{equation}
where $m_h^{+2(3)\sigma}=277(350)$ GeV.

In the plots we display the corresponding curves. Thus one sees that
the SM upper limit on the Higgs boson mass is now replaced by
restrictions in the $m_H$-$\cos\alpha$ plane which is consistent with
having $\Lambda$ of the order of few TeV, as seen in
Fig.~\ref{fig:cutoff}.

In the case in which the lightest CP-even Higgs boson is purely
doublet, that is when $\cos\alpha=0$, the heavier CP-even Higgs boson
mass $m_H$ is unconstrained, but we do not have an improvement with
respect to the SM as suggested by \eq{eq:limit-1}.  On the contrary,
in the opposite limit when the heaviest CP-even Higgs boson is purely
doublet, it is constrained as the Higgs boson in the SM, since in this
case it acts as ``the'' effective Higgs scalar.  As the value of
$\cos\alpha$ is varied the EWPT constraint correspondingly weakens.

\begin{figure}[ht] \centering
 \includegraphics[clip,height=8cm,width=\linewidth]{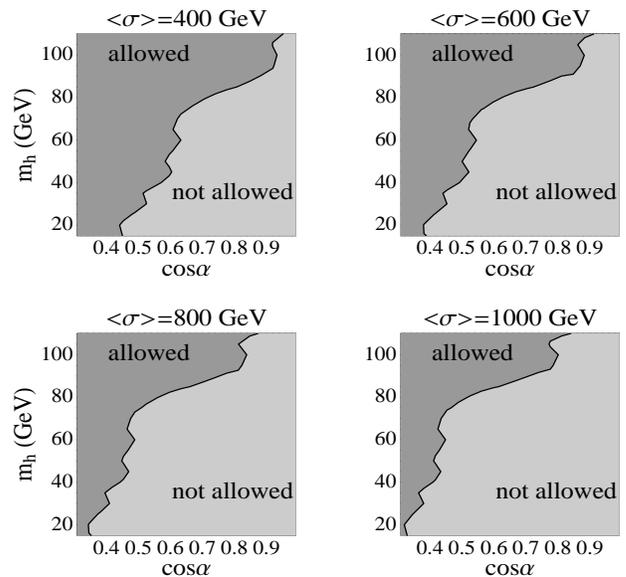}
    \caption{\label{fig:lowboundmh}
Regions allowed by LEP-II searches }
\end{figure}

Searches for invisibly decaying Higgs bosons using the LEP-II data
have been preformed by the LEP Collaborations.
For the channel $e^+ e^- \to Z h \to Z b\bar{b}$ the final state is
expressed in terms of the SM hZ cross section through
\begin{eqnarray}
\label{eq:defsigma}
\sigma_{hZ \to b\bar{b}Z} &=&\sigma_{h Z}^{SM}\times R_{h Z}\times BR(h \to b \bar{b}) \nn\\
&&\sigma_{hZ}^{SM}\times C^2_{Z(h \to  b\bar{b}) }\,, 
\end{eqnarray}
where $R_{hZ}$ is the suppression factor related to the coupling of
the Higgs boson to the gauge boson Z (i.e.  $R_{hZ}^{SM}=1$ and for
the model we have $R_{hZ}=\cos^2\alpha$). Here $BR(h \to b \bar{b})$
is the branching ratio of the channel $h \to b \bar{b}$ which in the
model is modified with respect to the SM both by the mixing angle
$\alpha$ and by the presence of the invisible Higgs boson decay into
the Goldstone boson J associated to the breaking of the global $U(1)$
symmetry, Eq.~(\ref{eq:jj}).  For example in
Ref.~\cite{Abdallah:2004wy} DELPHI gives upper bounds for the
coefficients $ C^2_{Z(h \to b\bar{b})}$ corresponding to a lightest
CP-even Higgs boson mass from $15$ GeV up to $100$ GeV. We have
analysed the regions of the mixing angle $\alpha$ and of the
parameters $v_\sigma$ and $m_h$ which are currently allowed by the
LEP-II searches.  The results are illustrated in
Fig.~\ref{fig:lowboundmh}.
One sees that as $\cos\alpha \to 1$ $h$ must obey the LEP-II SM Higgs
boson limit while, in the opposite limit where it becomes fully
singlet, the limit quickly deteriorates.

There are many model realizations sharing the same simplest Higgs
scalar potential in Eq.~(\ref{eq:pot}). Models differ depending on the
details of the couplings relevant for neutrino masses. We assume that
we have a non-supersymmetric model of neutrino
masses~\footnote{Obviously supersymmetry may always be added but, in
  the ``worse-case'' spirit of Ref.~\cite{Barbieri:2005kf} which we
  are adopting, that would be an over-kill. }.  A simple tree-level
example is the ``inverse seesaw'' model introduced in
\cite{gonzalez-garcia:1989rw}, described by
\begin{equation}
{\cal L}_Y = Y_{ij} {\nu^c}_{iL} \ell_{jL} \Phi + 
M_{ij} {\nu^c}_{iL} S_j 
+ \lambda_{ij}  S_i S_j \sigma  
\end{equation}
In addition to the \321 singlet ``right-handed'' neutrinos it contains
gauge singlet leptons $S_i$ which may arise in some string
models~\cite{mohapatra:1986bd}. No other terms are allowed by the
symmetry, e.~g., a direct Majorana mass term for the singlet fields
$S_i$ is forbidden by lepton number and arises only due to the singlet
scalar vev $\vev\sigma$.
This will then give  a $9 \times 9$ neutrino mass matrix, in the
basis ($\nu_i$, $\nu^{c}_i$, $S_i$):
\begin{equation}
\label{eq:nu-mass-mat}
M_{\nu}=\left(
\begin{array}{ccc}
0  &Y_\nu v_\Phi & 0  \\
Y^{T}_\nu v_\Phi & 0 & M \\
0 & M^T & \mu \\
\end{array}
\right)
\end{equation}
so that the effective left-handed light neutrino mass matrix is
\begin{equation}
\label{eqn:lightNu}
    m_\nu = {m_D^T M^{T}}^{-1} \mu M^{-1} m_D .
\end{equation}
In the limit $\mu \to 0$ lepton number is restored and
neutrino masses vanish. Thus its smallness is ``natural'' and
$\vev\sigma$ can easily be of the order of TeV or less.

We have presented only the very simplest example of a class of
low-scale neutrino mass models which can be made natural up to few TeV
or so, in the absence of supersymmetry or other stabilizing
mechanisms. These are a viable and attractive alternative to the
seesaw mechanism. As explained in \cite{Joshipura:1992hp} and briefly
reviewed in \cite{Valle:2006vb}, there are many models with
spontaneous breaking of lepton number at low scale, in which case the
majoron is present. All of these lead to the generic signal in
Eq.~(\ref{eq:jj}).  Thus from this point of view the possibility of
invisibly decays Higgs bosons must be taken seriously from the point
of view of future colliders, LHC and ILC.
Additional signals associated, say, to extended Higgs sector and
charged scalars may also exist. These survive even if lepton number
breaking is taken as explicit.
A more extended investigation of these schemes will be presented
elsewhere~\cite{prepa}.

\def\baselinestretch{1}%
Work supported by MEC grants FPA2005-01269 and BFM2002-00345, by EC
Contracts RTN network MRTN-CT-2004-503369.  F.~B.  is supported by a
MEC postdoctoral grant.

 \def\baselinestretch{1}%


\end{document}